\documentclass[mathleft,fleqn,%
]{an}
\usepackage{graphicx}
\usepackage[varg]{txfonts}
\usepackage{natbib}
\bibpunct{(}{)}{;}{a}{}{,}
\definecolor{com}{rgb}{0.698039, 0.0941176, 0.133333}

\begin{document}


\Pagespan{1}{}
\Yearpublication{2016}
\Yearsubmission{2016}
\Month{0}
\Volume{999}
\Issue{0}
\DOI{asna.201600000}

\title{Spectropolarimetric observations of an arch filament system\\
    with the GREGOR solar telescope}

\author{%
    H.\ Balthasar\inst{1}\fnmsep\thanks{Corresponding author: 
    {hbalthasar@aip.de}}
\and 
    P.\ G\"om\"ory\inst{2}
\and 
    S.J.\ Gonz{\'a}lez Manrique\inst{1,3}
\and 
    C.\ Kuckein\inst{1}
\and
    J.\ Kavka\inst{2}
\and 
    A.\ Ku{\v c}era\inst{2}
\and\\ 
    P.\ Schwartz\inst{2}
\and 
    R.\ Va{\v s}kov{\'a}\inst{2}
\and 
    T.\ Berkefeld \inst{4}
\and 
    M.\ Collados Vera\inst{5}
\and 
    C.\ Denker\inst{1}
\and 
    A. Feller\inst{6}
\and    
    A.\ Hofmann\inst{1}
\and \\
    A.\ Lagg\inst{6}
\and 
    H.\ Nicklas\inst{7}
\and
    D. Orozco Su{\'a}rez\inst{8}
\and
    A. Pastor Yabar\inst{5}
\and
    R. Rezaei\inst{5}
\and
    R. Schlichenmaier\inst{4}
\and \\   
    D.\ Schmidt\inst{9}
\and 
    W.\ Schmidt\inst{4}
\and 
    M.\ Sigwarth\inst{4}
\and 
    M.\ Sobotka\inst{10}
\and 
    S.K.\ Solanki\inst{6,11}
\and 
    D.\ Soltau\inst{4}
\and 
    J.\ Staude\inst{1}
\and \\ 
   K.G.\ Strassmeier\inst{1}
\and
    R.\ Volkmer\inst{4}
\and 
    O.\ von der L\"uhe \inst{4} 
\and 
    T.\ Waldmann\inst{4}}

\titlerunning{Spectropolarimetric observations of an arch filament system}

\authorrunning{H.\ Balthasar et al.}

\institute{
    Leibniz-Institut f\"ur Astrophysik Potsdam (AIP), 
    An der Sternwarte 16, 
    14482 Potsdam, Germany
\and 
    Astronomical Institute of the Slovak Academy of Sciences, 
    Tatransk\'a Lomnica, Slovak Republic
\and 
    Universi\"at Potsdam, 
    Institut f\"ur Physik und Astronomie, 
    Karl-Liebknechtstra{\ss}e 24/25, 14476 Potsdam-Golm, Germany
\and
    Kiepenheuer-Institut f\"ur Sonnenphysik, 
    Sch\"oneckstra{\ss}e 6, 79104 
    Freiburg, Germany
\and 
    Instituto de Astrof\'{\i}sica de Canarias, 
    C/ V\'{\i}a L\'actea, s/n, 38205 La Laguna (Tenerife), Spain
\and
    Max-Planck-Institut f\"ur Sonnensystemforschung, 
    Justus-von-Liebig Weg 3, 
    37077 G\"ottingen, Germany
\and
    Georg-August-Universit\"at G\"ottingen, 
    Institut f\"ur Astrophysik, 
    Friedrich-Hund-Platz 1, 37077 G\"ottingen, Germany
\and
    Instituto de Astrof{\'\i}sica de Andaluc{\'\i}a - CSIC, 
    Glorieta de la Astronom{\'\i}a, s/n, 18008 Granada, Spain
\and
    National Solar Observatory, 
    3010 Coronal Loop,
    Sunspot, NM 88349, U.S.A. 
\and
    Astronomical Institute, Academy of Sciences of the Czech Republic, 
    Fri{\v c}ova 258, 
    Ond{\v r}ejov, Czech Republic
\and
    Kyung Hee University,
    Yongin, Gyeonggi-Do, 446 701 Republic of Korea
    }
\received{\today}
\accepted{XXXX}
\publonline{XXXX}

\abstract{Arch filament systems occur in active 
sunspot groups, where a fibril 
structure connects areas of opposite magnetic polarity, in contrast to active 
region filaments that follow the polarity inversion line. We used the GREGOR 
Infrared Spectrograph (GRIS) to obtain the full Stokes vector in the spectral 
lines Si\,\textsc{i} $\lambda$1082.7\,nm, He\,\textsc{i} $\lambda$1083.0\,nm, 
and Ca\,\textsc{i} $\lambda$1083.9\,nm. We focus on the near-infrared calcium 
line to investigate the photospheric magnetic field and velocities, 
and use the line core intensities and velocities of the helium line to study the 
chromospheric plasma. 
The individual fibrils of the arch filament system connect the sunspot with 
patches of magnetic polarity opposite to that of the spot. 
These patches do not necessarily coincide with pores, where the magnetic field is 
strongest. Instead, areas are preferred not far from the polarity inversion line. 
These areas exhibit photospheric downflows of moderate velocity, but 
significantly higher downflows of up to 30\,km\,s$^{-1}$ in the chromospheric 
helium line. 
Our findings can be explained with new emerging flux where the matter flows downward 
along the fieldlines of rising flux tubes, in agreement with earlier results.
}
\maketitle


\section{Introduction}

The structure of solar filaments depends on the magnetic field at their 
photospheric footpoints. Thus, to understand the processes forming a filament, 
a good knowledge of the underlying magnetic vector field is mandatory. In active 
regions, one has to distinguish between active region filaments and arch 
filament systems (AFS). Active region filaments follow the neutral or polarity 
inversion line (PIL), while structures in AFS cross the PIL. 
Commonly, active region filaments are related to magnetic shearing at the PIL, and 
AFSs appear where the emergence of new flux is still ongoing.
The physical processes that lead in both cases to dark structures in chromospheric 
lines need to be investigated in more detail. It is generally assumed that the dark 
structures of an AFS represent the emerging fluxtubes. In this work we investigate such an AFS.
For active region filaments we refer to recent works of \cite{cit:kuck12mag},
\cite{cit:kuck12vel}, \cite{cit:sasso2014} and \cite{cit:schwartz2016} 
and references therein.

\begin{figure}[t]
\includegraphics[width=\columnwidth]{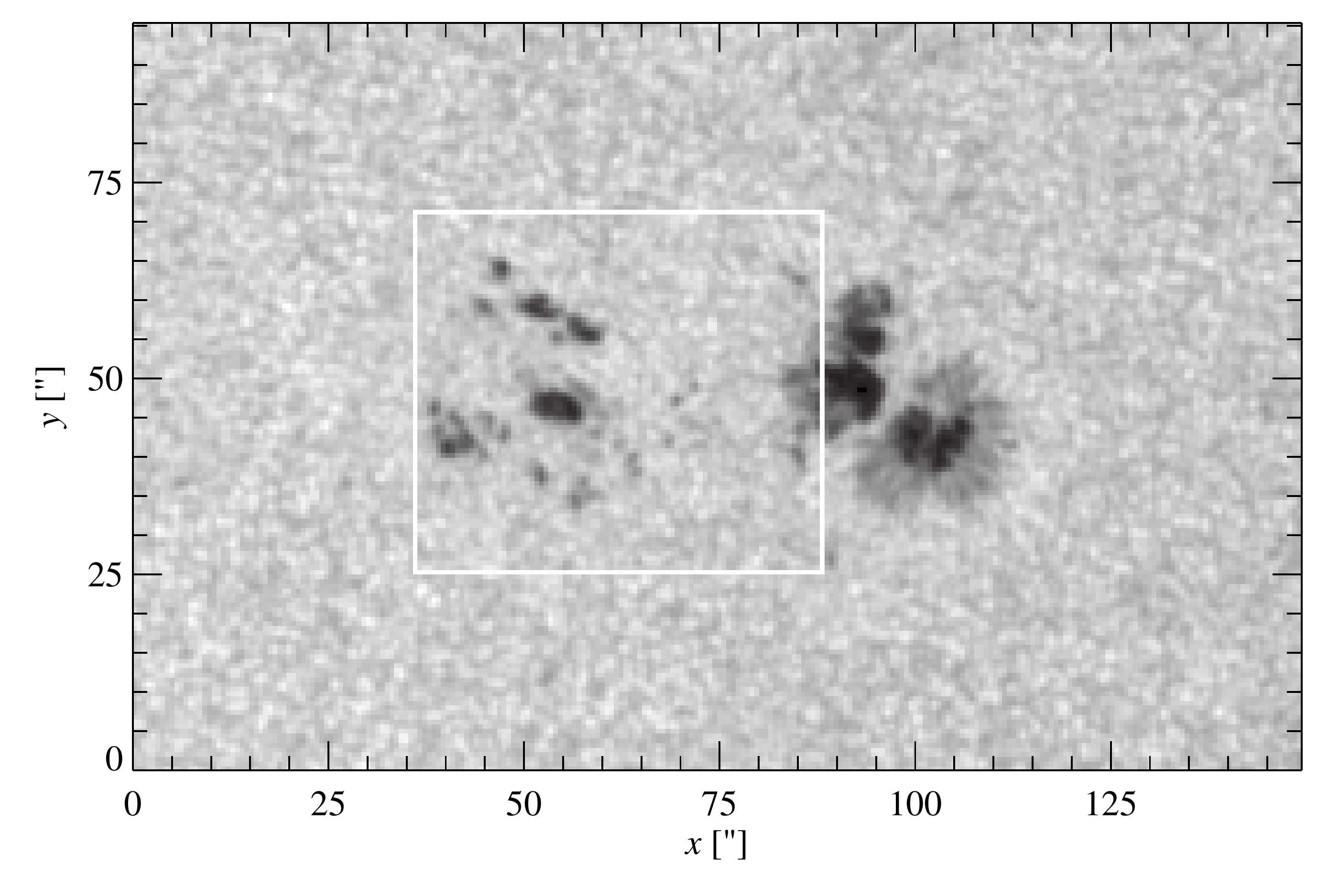}
\caption{Intensity image taken at 08:16\,UT by HMI. The white box outlines the 
area investigated in this work.
}
\label{fig_hmi}
\end{figure}

\begin{figure*}[t]
\includegraphics[width=\textwidth]{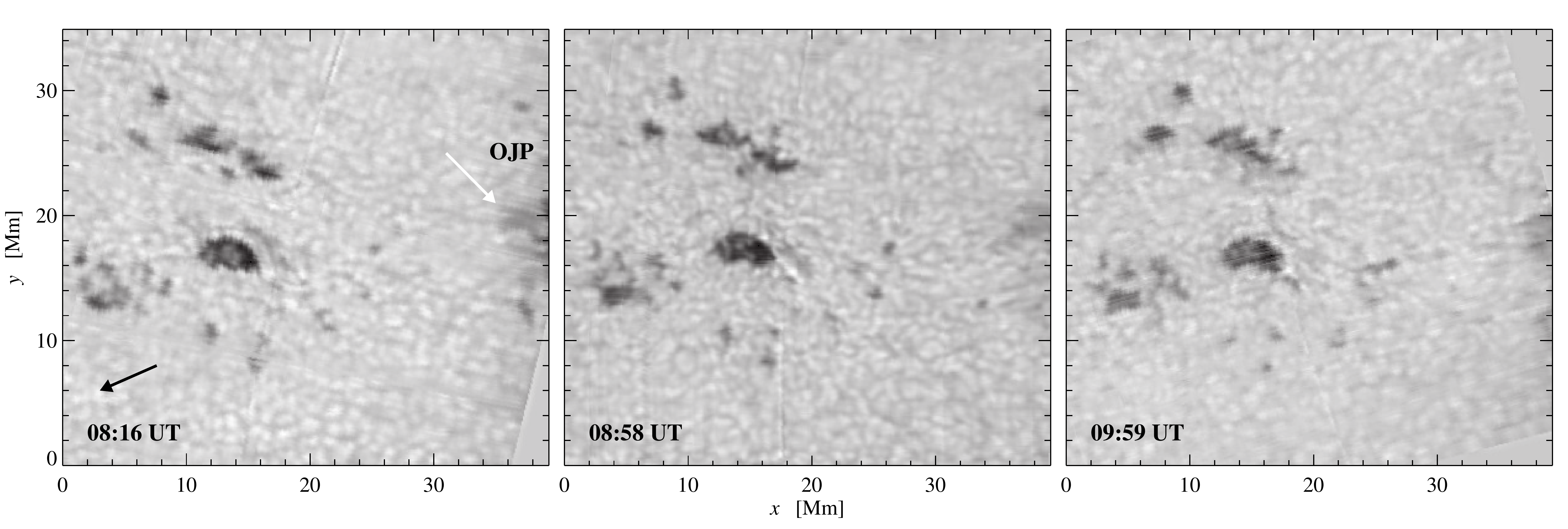}
\caption{Temporal evolution of the slit-reconstructed continuum images. 
    We apply an unsharp masking to enhance the displayed contrast. 
    The black arrow points towards disk center and the white arrow to an outjutting
    part of the penumbra (OJP) of the nearby spot.} 
\label{fig_int}
\end{figure*}

\cite{cit:howardharvey} distinguished between filaments and fibrils, where 
filaments lie parallel to iso-contour lines of the magnetic field and fibrils 
traverse them perpendicularly. \cite{cit:martres66} observed fibrils (``traces 
filamenteuses'') that connect areas of opposite polarity crossing the PIL. 
Systems of fibrils were described by \cite{cit:bruzek67}, and he suggested to 
call them AFS. They appear in the inter spot area and connect spots of opposite 
polarities and cross the PIL. Strong downflows of about 50\,km\,s$^{-1}$ occur 
at the footpoints of arch filaments \citep{cit:bruzek69}. 
The dynamic evolution of an AFS was investigated by \cite{cit:spadaro}, who found
upward motions in central parts of the fibrils and downward flows at their ends.
Recently, 
\cite{cit:vargas} used Hinode data to study AFS related to granular scale flux 
emergence in an active region. \cite{cit:grigoreva} considered the formation of 
AFS in an active region as the onset of newly emerging flux before a strong 
flare occurred one day later. \cite{cit:ma} investigate an AFS in H$\alpha$, and 
they use magnetograms from the Helioseismic and Magnetic Imager (HMI) onboard the 
Solar Dynamics Observatory (SDO). In their case, the arch filaments are related to 
moving magnetic features of opposite polarity to that of the main sunspot and 
the arch filaments are controlled by the moving magnetic features.
\cite{cit:saminat} and \cite{cit:lagg07} observed an AFS in the infrared helium line
at 1083\,nm and determined its magnetic structure.
The chromospheric magnetic field of another AFS is investigated by \cite{cit:xu10}. 
In the zone of flux emergence they find a magnetic field strength of 300\,G, 
and the magnetic field is almost horizontal. At the edge of the emerging zone, the
field is more vertical and stronger with a field strength of up to 850\,G.
Supersonic downflows occur at both footpoints of a fibril (or loop as the authors 
call the structures).

In the present work, we investigate an AFS, and we will concentrate on the 
magnetic field and the Doppler velocities in the lower photosphere. According to 
\cite{cit:howardharvey} we will use the term ``fibril'' instead of ``arch 
filament'' for the single structures of the AFS in this work. Results of the 
upper photosphere and chromosphere will be topic of another article, except for 
a velocity determination from the infrared helium triplet line at $\lambda$1083\,nm.

\begin{figure*}[t]
\includegraphics[width=\textwidth]{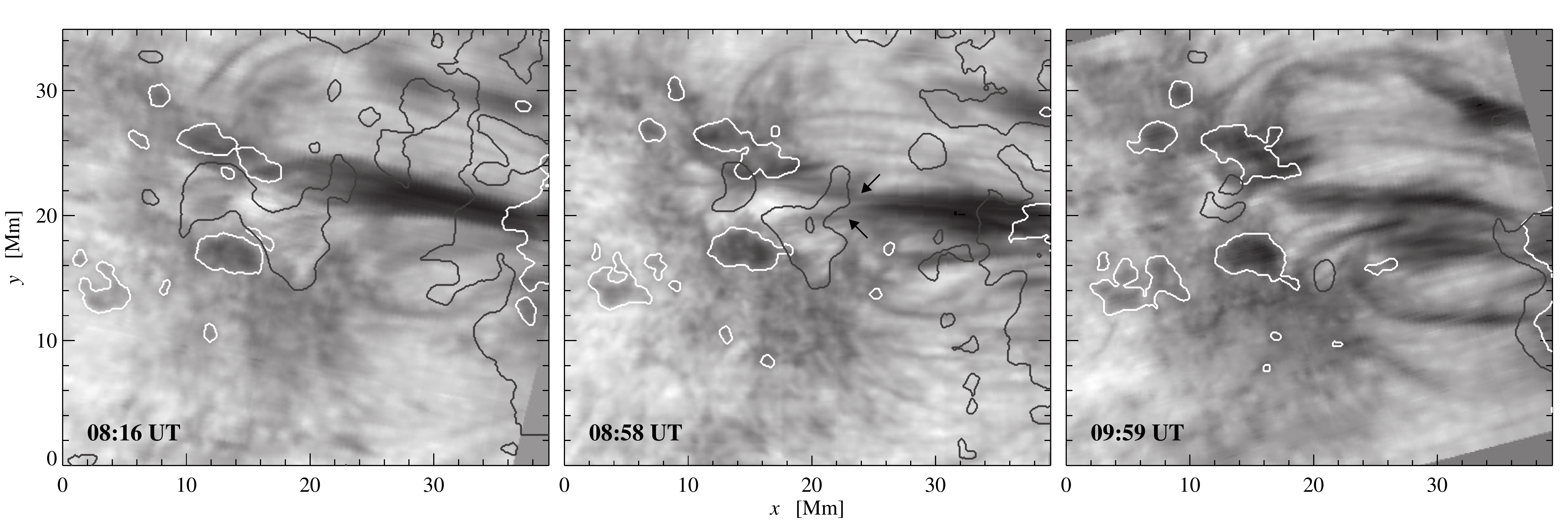}
\caption{Temporal evolution of the slit-reconstructed helium line core 
    intensity. White contours mark the dark photospheric intensity 
    structures as seen in Fig.\,\ref{fig_int}, and the black ones indicate the PILs. 
    The small arrows in the mid panel indicate the splitting of the 
    main fibril.}
\label{fig_inthe}
\end{figure*}


\section{Observations}

A small AFS in active region NOAA~12353 was observed on 2015 May~24 at the 
1.5-meter GREGOR solar telescope \citep{cit:gregor, cit:kneer, cit:denker12}. 
The whole group is depicted in Fig.\,\ref{fig_hmi}. 
It was located at $x = 380\arcsec$ and $y = 140\arcsec$, 
corresponding to cos$\vartheta$ = 0.904. The 
image was stabilized and corrected in real-time by the GREGOR Adaptive Optics 
System \citep[GAOS,][]{cit:gaos}. The AO system locked on the central pore in 
the marked subfield of Fig.\,\ref{fig_hmi}.

We used the GREGOR Infrared Spectrograph \citep[GRIS,][]{cit:gris} in 
spectropolarimetric mode.
More details about the polarimeter are presented in \cite{cit:tip2}.
The spectral range was 1082.4\,--\,1084.2\,nm which 
covers the silicon line $\lambda$1082.7\,nm from the upper photosphere, the 
chromospheric helium triplet  $\lambda$1083.0\,nm and the calcium line 
$\lambda$1083.9\,nm from the deep photosphere. The silicon and the calcium lines 
are both Zeeman triplets with an effective Land\'e-factor of $g_\mathrm{eff} 
=1.5$. The spectral dispersion was 1.810\,pm per pixel, and the spectral resolution 
element is 5.7\,pm assuming the measured resolution power given by \cite{cit:gris}. 
\cite{cit:lagg16} re-determined the spectral resolution power in a lower order to 110\,000
very close to the value of \cite{cit:gris}.
We scanned five times across the 
region with the slit scanner in the time interval 08:06\,--\,10:08\,UT. The step 
size was 0\farcs 135, and  a scan consisted of 400 steps, except for the first 
scan when we carried out 450 steps. We covered roughly 60$^{\prime\prime}$ 
along the slit. The single exposure time was 100\,ms, and we accumulated four 
cycles through the different states of the ferro-electric liquid crystals. A 
scan of 400 steps thus takes about 18\,min.

\begin{figure*}
\includegraphics[width=\textwidth]{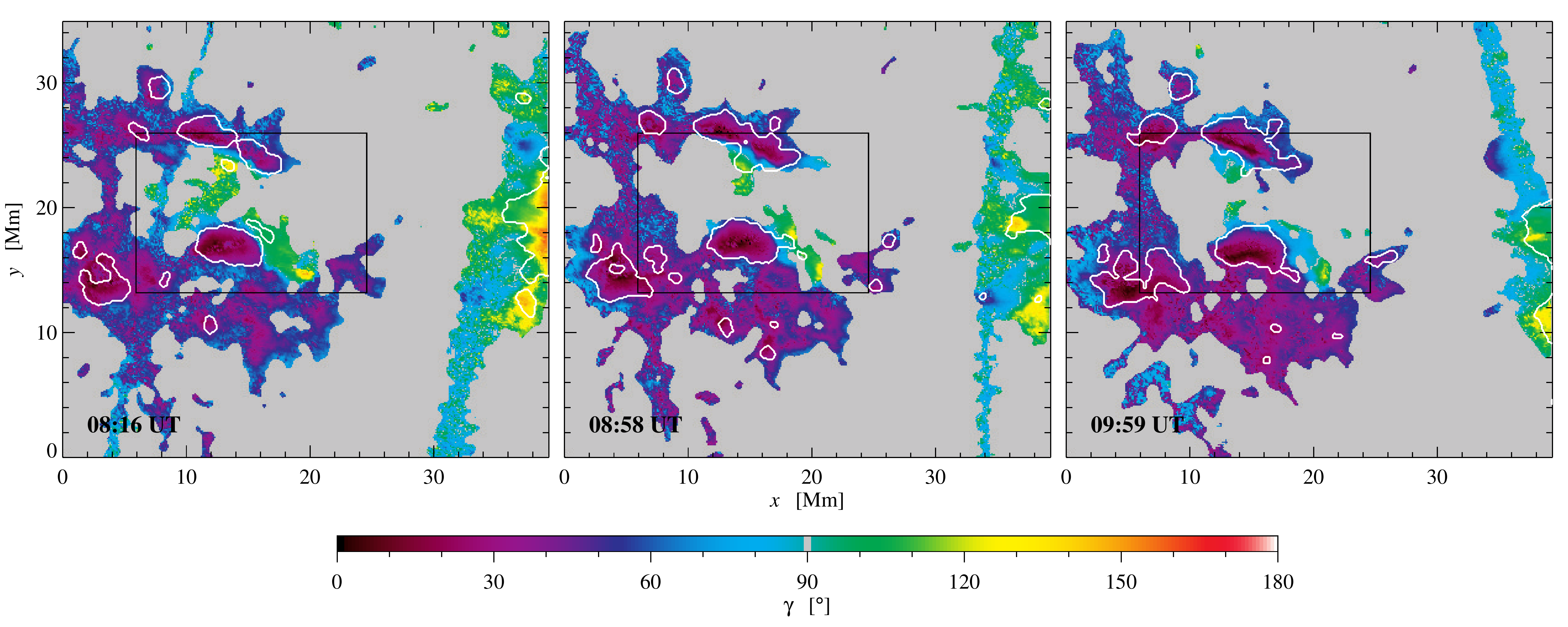}
\caption{The inclination of the  magnetic field strength for three different 
    scans. White contours indicate the dark photospheric structures.
    The rectangular boxes indicate the area over which we integrated the
    negative magnetic flux, see text. In the gray areas, the polarization signal is 
    below the significance threshold. }
\label{fig_gamm}
\end{figure*}

\begin{figure*}
\includegraphics[width=\textwidth]{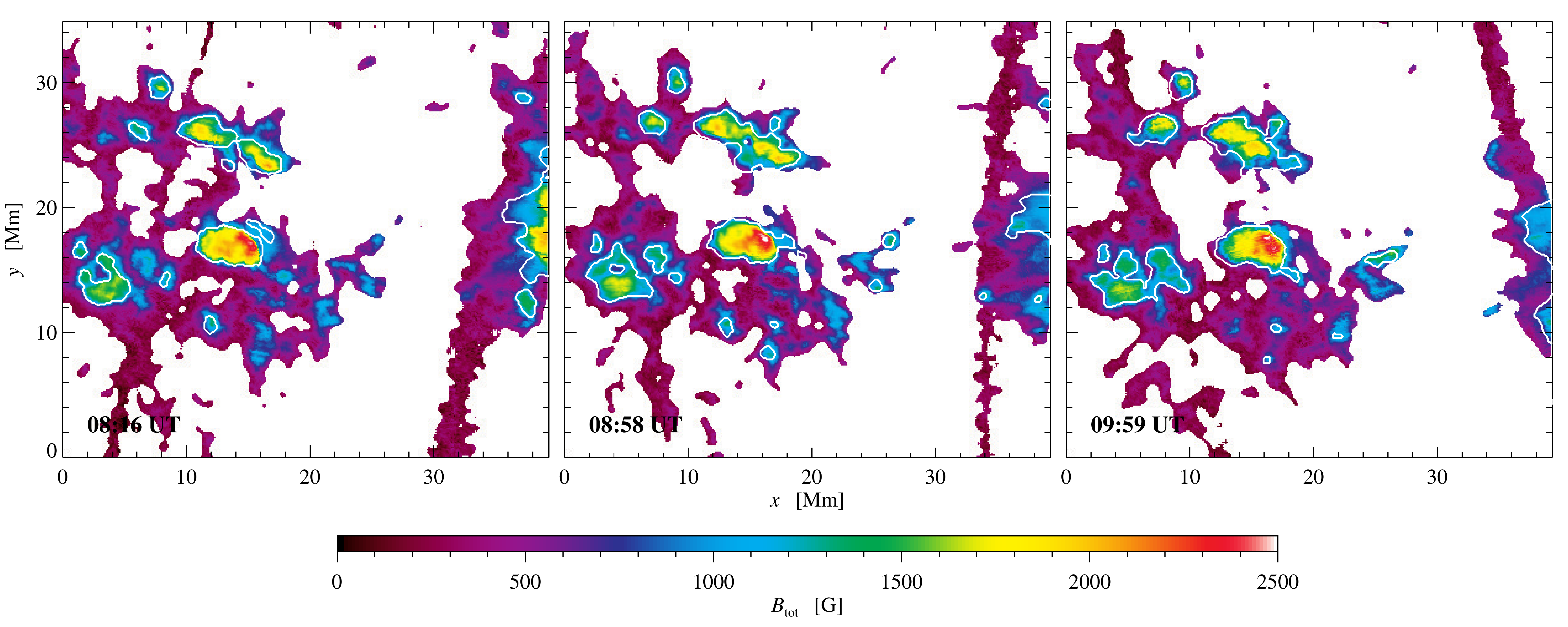}
\caption{The total magnetic field strength for three different scans. White 
    contours indicate the dark photospheric structures.
    In the white areas, the polarization signal is 
    below the significance threshold.}
\label{fig_btot}
\end{figure*}


\section{Data reduction}

The data were corrected using dark and flat-field frames. The polarimetric 
calibration was carried out with measurements taken with the GREGOR Polarimetric 
calibration Unit \citep[GPU,][]{cit:gpu}. The demodulation of the measurements 
follows the description by \cite{cit:mcv}. 
A crosstalk correction is performed similar as described by 
\cite{cit:colschlich} for the VTT, but for the GRIS-data, only the influence of 
Stokes $I$ on $Q$, $U$ and $V$ is corrected to force the continuum in the 
polarization parameters to zero. No correction for the crosstalk between linear 
and circular polarization is applied.   
To derive the magnetic vector field, 
we use the code Stokes Inversion based on Response functions 
\citep[SIR,][]{cit:sir}. Since the calcium line is rather weak, it forms in a 
narrow atmospheric height layer, and we can restrict the calculation to one node 
for Doppler velocity, magnetic field strength, inclination, and azimuth, and we 
keep these values height-independent. Only for the temperature, we do allow three 
nodes. We assumed a single atmospheric component. 
The dispersed 
straylight was kept fixed at a level of 2\%. This value seems to be rather low, 
but we found the fits to be more reliable than with a significantly higher amount of 
straylight. The true amount of straylight is probably higher, but in our case   
the assumption of unpolarized light from the non-magnetic quiet sun, as only foreseen in the 
SIR-code, is insufficient.

Photospheric Doppler shifts suffer from temporal spectrograph drifts. We assume 
that the wavelength of a nearby telluric line caused by water vapor does not 
change, so that we can use it to correct for the spectrograph drifts. We 
determine its wavelength by applying a polynomial fit through the line core. The 
drifts correspond to several 100\,m\,s$^{-1}$ during a single scan. 

All maps 
shown in this article are compensated for the differential image rotation
(for details see Appendix~\ref{sec:dirc}). The magnetic vector is transferred to 
the local frame of reference.
Then, all images are rotated as a whole to have the $y$-axis parallel to the 
solar North-South direction, and the geometrical foreshortening has been 
corrected. Finally, we cut out the common area as shown in Fig\,\ref{fig_int}.


\section{Results}

The morphology and dynamics of AFS are closely linked to the photospheric and
chromospheric flows and magnetic fields. 
A display of three slit-reconstructed continuum images is shown in 
Fig.\,\ref{fig_int}. 
Since the evolution in the area was not too fast, 
we show only three of the five scans in this and the following figures.
We scanned an area containing several pores of the 
following part of the group, while the main spot was just outside of this area. 
The AFS 
connected the main spot just outside at the right of Fig.\,\ref{fig_int}
with following pores of opposite magnetic polarity (see Fig.\,\ref{fig_inthe}). 
The fibrils were more or less perpendicular to the PIL. 
In the first scan, the AFS is rather compact and connects an 
outjutting (protruding) part of 
the penumbra (OJP), indicated by the white arrow in Fig.\,\ref{fig_int}
with the 
merging pores at $x = 10 - 20$\,Mm and $y = 26$\,Mm. Then the AFS becomes weaker on 
the side towards the pores, and we see two branches, marked by little 
arrows in the mid panel of Fig.\,\ref{fig_inthe}. 
The stronger branch now 
reaches the central pore at $x = 20$\,Mm and $y = 18$\,Mm. In the last scan the 
AFS separates into several substructures, and the OJP is no longer in the field of view. 
In addition, some arc structures, 
barely visible before, become more pronounced. The central pore did not change 
very much, but the two pores above were merging during our observing period.
In the following, we present the 
results from near-infrared spectropolarimetry.


\subsection{Magnetic field}

The larger pores have a positive magnetic polarity corresponding to small 
values for the magnetic inclination as seen in Fig.\,\ref{fig_gamm}. The leading 
spot has negative polarity, and there is negative polarity between the pores 
(inclination larger than 90$^\circ$). The total magnetic field strength is 
displayed in Fig.\,\ref{fig_btot}. The central pore is the strongest one and 
contains fields of up to 2500\,G. The two merging pores have cores with field 
strengths of about 2000\,G in the first scan. Other pores have 
1600\,--\,1800\,G. Below the AFS, we encounter magnetic fields with 
200\,--\,400\,G, and the vertical component is less than 100\,G with varying 
sign. The magnetic field at the edge of the penumbra amounts to 
900\,--\,1000\,G. In the OJP, the vertical component of the magnetic field 
amounts to values between  $-300$ and $-200$\,G, as expected for a spot with negative magnetic 
polarity. At 08:16\,UT, the magnetic field in the OJP is almost horizontal (~100$^\circ$),
but at 08:58\,UT, it is more vertical (140$^\circ$\,--\,150$^\circ$).
In contrast, the nearby areas show a positive component of several 
hundred Gauss.

We estimate the positive magnetic flux in the pores and their surroundings and 
the flux of opposite sign in the intermediate area between the pores by 
integrating the vertical component of the magnetic field. The negative flux between the
pores is integrated in the rectangular subfield with the coordinates 5.9\,Mm $< x <$ 24.6\,Mm
and 13.2\,Mm $< y <$ 26.0\,Mm, which is indicated by the rectangles in Fig.\,\ref{fig_gamm}.
Both the area 
containing negative flux and the magnetic flux of negative sign decreases with 
time. The value of positive flux is increasing with time, but the errors are quite large.
If real, this increase is much 
larger than the decaying negative flux. Values are given in 
Table\,\ref{tab:flux}. 

\begin{table}[t]
\caption{Magnetic flux in pores and intermediate area.}
\label{tab:flux}
\begin{tabular}{cccc}\hline
Scan  & Time & positive flux  & negative flux\rule[0.0mm]{0mm}{3.5mm} \\ 
      &      & ($10^{13}$ Wb) & ($10^{11}$ Wb)\rule[-1.5mm]{0mm}{3mm} \\
\hline
1 & 08:06\,UT & 1.5 $\pm$ 0.5 &  $-5.2$ $\pm$ 2.6\rule[0.0mm]{0mm}{3.5mm}\\
2 & 08:27\,UT & 1.2 $\pm$ 0.4 &  $-3.2$ $\pm$ 2.2\\ 
3 & 08:58\,UT & 1.7 $\pm$ 0.5 &  $-2.8$ $\pm$ 1.3\\
4 & 09:36\,UT & 1.8 $\pm$ 0.5 &  $-0.8$ $\pm$ 0.9\\
5 & 09:59\,UT & 1.8 $\pm$ 0.6 &  $-0.6$ $\pm$ 0.8\rule[-1.5mm]{0mm}{3mm}\\
\hline
\end{tabular}
\end{table}

\begin{figure*}
\includegraphics[width=\textwidth]{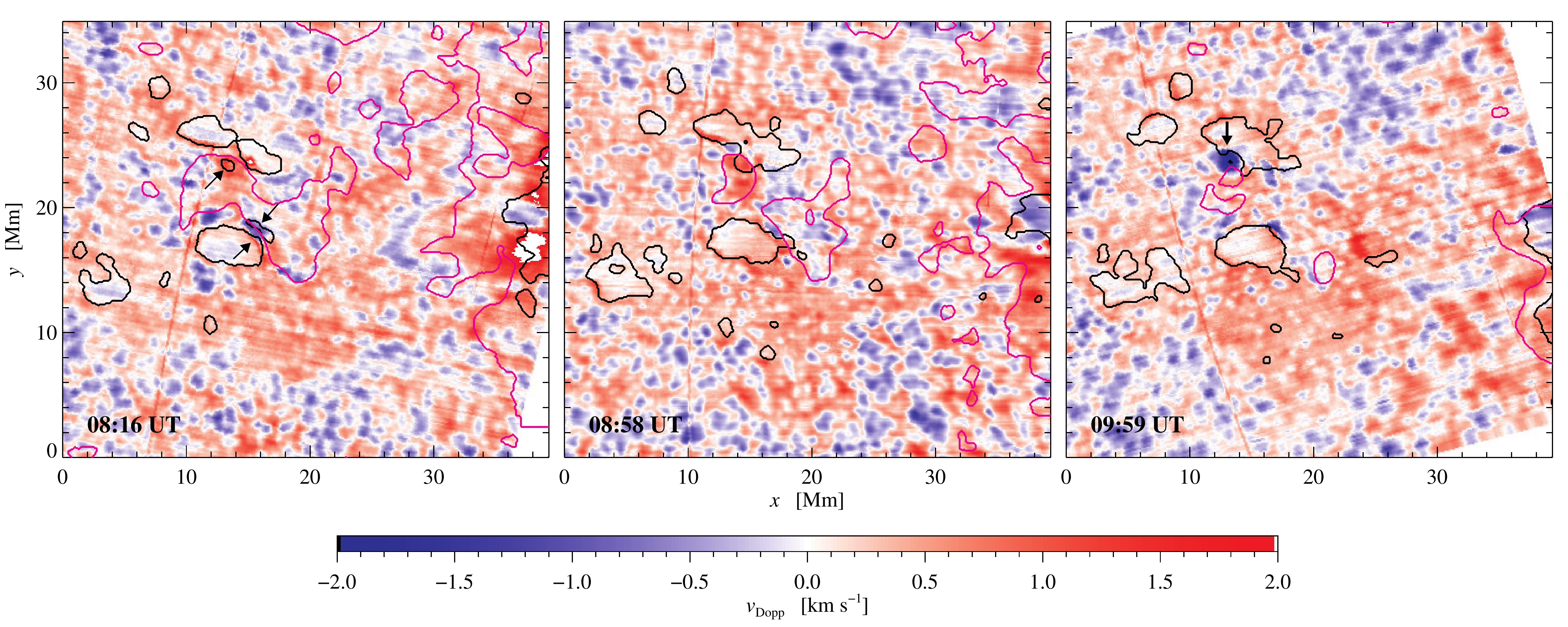}
\caption{Doppler velocities determined from the photospheric calcium line. The 
    black contours mark the dark photospheric structures and the 
    magenta contours indicate the PILs. Values larger than 2\,km\,s$^{-1}$ are 
    clipped. Little arrows indicate special patches discussed in the text.}
\label{fig_velo}
\end{figure*}


\subsection{Photospheric Doppler velocities}

Photospheric Doppler velocities are displayed in Fig.\,\ref{fig_velo}. Inside 
the central pore, we encounter very low velocities, but at the edge towards the 
opposite magnetic polarity, we see a patch of redshifts and just outside at the 
PIL blueshifts, which might indicate a horizontal flow roll.
These patches are marked by little arrows in Fig.\,\ref{fig_velo}, as well as 
the following one.
At the edge of the two merging pores, where the 
dark fibrils end, the line is redshifted during the scans at 08:16\,UT and at 
08:58\,UT (about 2\,km\,s$^{-1}$ in the beginning). In the last scan at 
09:59\,UT, we find a blueshift of about $-2$\,km\,s$^{-1}$ instead of the 
previously observed downflow. The OJP
exhibits 
mainly a blueshift of about $-0.5$\,km\,s$^{-1}$. A blueshift here is in 
agreement with the expected Evershed effect, because the disk center is towards 
the left. However, on both sides of the 
OJP, we see strong redshifts up to 5\,km\,s$^{-1}$ inside the penumbra. 
A similar configuration was described by \cite{cit:schliche273} and 
\cite{cit:schliche_hinode}, in that case 
it was observed during the formation of the penumbra.
In the intermediate area between the 
main spot and the pores, we find dominantly redshifts in the order of 
0.4\,--\,1\,km\,s$^{-1}$.


\subsection{Chromospheric Doppler velocities}

\begin{figure*}
\includegraphics[width=\textwidth]{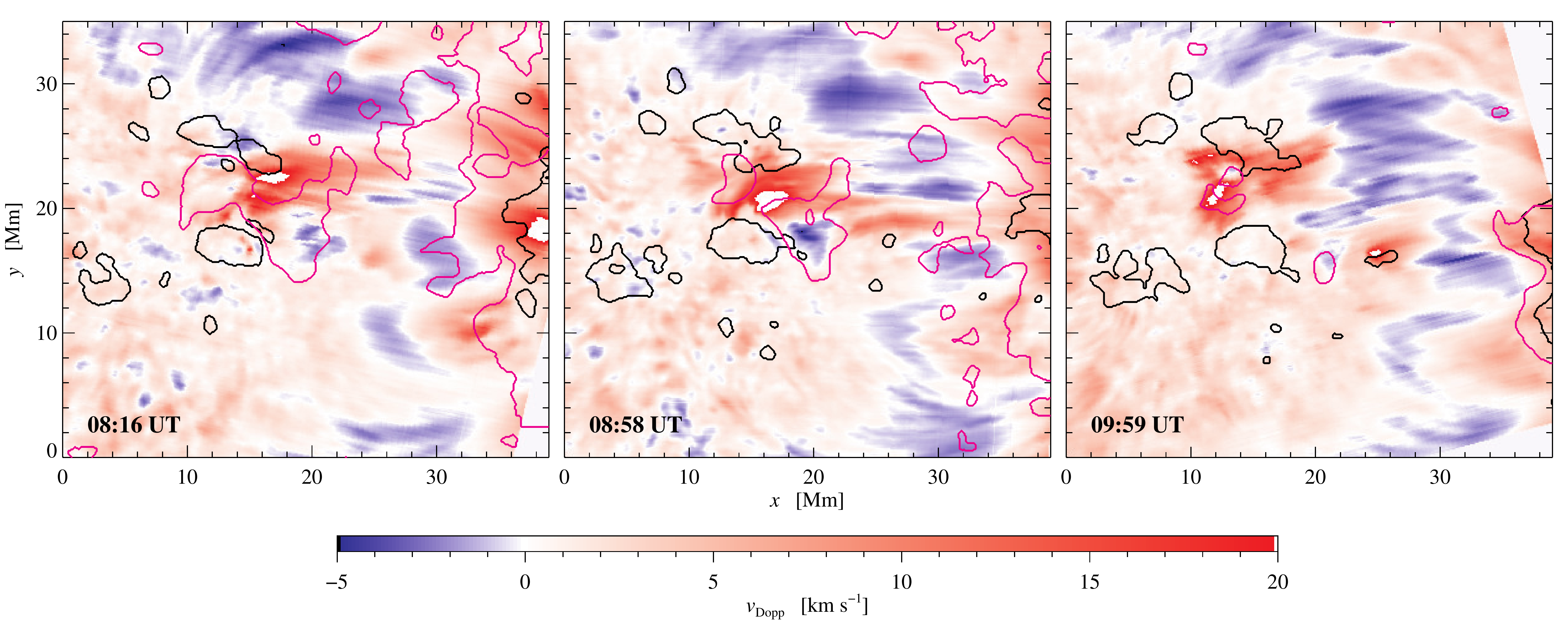}
\caption{Doppler velocities determined from the helium line. The velocities are 
    limited to 20\,km\,s$^{-1}$, i.e., in the white patches they reach up to 
    30\,km\,s$^{-1}$. Black contours indicate the dark photospheric 
    structures and magenta ones the photospheric PILs.}
\label{fig_velhel}
\end{figure*}

Chromospheric Doppler velocities are determined from the helium line by applying 
the method introduced by \cite{cit:sergio}. The velocities are calibrated on an 
absolute scale following the steps in Appendices A and B of \cite{cit:kuck12vel}. 
Here, we present the results for a 
one-component line fit. Upflows (blueshifts) amount to up to 5\,km\,s$^{-1}$ and 
appear mainly with an arch-like structure between the main spot and the opposite 
polarity pores. In general, these blueshifts do not coincide with the PIL of the 
deep photosphere. Downflows (redshifts) reach values up to 30\,km\,s$^{-1}$. In 
the beginning at 08:16\,UT, they occur next to the end of the most pronounced 
arch at the merging pores. We also find a redshift above the penumbra, while 
there is a blueshift in the OJP, in the photosphere caused by 
the Evershed effect (see Fig.\,\ref{fig_velo}). Later we encounter the maximum 
values between the central pore and the merging ones hitting the polarity 
inversion lines.
The results are displayed in Fig.\,\ref{fig_velhel}. A similar distribution of strong redshifts
(up to 15\,km\,s$^{-1}$) at the footpoints of magnetic loops and moderate 
blueshifts (up to -5\,km\,s$^{-1}$) in their central parts was observed by \cite{cit:lagg07}. 


\section{Discussion}

We examine an AFS, where the fibrils cross the photospheric PIL, and it 
is likely that the fibrils connect areas of different polarities, in contrast to active 
region filaments that follow the PIL. AFSs are related to new emerging bipolar magnetic flux. 
We conjecture that the physics behind the fibrils 
described by \cite{cit:howardharvey} and the arch filaments observed by 
\cite{cit:bruzek67} is very similar, and that the difference lies in their 
appearance in H$\alpha$. Unfortunately, we do not have  H$\alpha$ data taken in 
parallel to our observations. 

In the beginning of our observations at 08:16~UT, the dark fibrils connect 
apparently the OJP, where we see the normal Evershed effect in 
photospheric layers, with the region next to the merging pores, where the matter 
is flowing downward. 
Next to the OJP, 
we encounter redshifts, probably downflows. In the helium 
line, we see downflows above the penumbra, but the strongest downflows appear 
at the other footpoints of the fibrils. These chromospheric downflows correspond 
probably to the photospheric downflows next to the OJP.
However, \cite{cit:schad} find that is rather difficult to connect the involved 
height layers, even although they performed an inversion of the helium line.
In the present state of our work, we do not know in which height layer the helium 
line is formed, perhaps so 
high that there is a big spatial gap to the photosphere.
The magnetic field in the OJP becomes 
less inclined with time, and at the end of our observations, the OJP is no longer 
in the field of view. It seems that this area plays an important role
for the understanding
of the fibrils that are dark in the helium core, but for sophisticated conclusions 
we would need the results from different atmospheric layers that we do not have 
already for this paper. The fibrils seem to end in most cases at 
patches with opposite polarity to that of the main spot, but at locations that 
are not very far from the PIL. These patches do not necessarily coincide with 
the dark pores, where the magnetic field is strongest.

Possibly, rising bipolar flux cancels the negative magnetic polarity between the 
pores. The positive magnetic flux in the pores remains more or less constant, 
but the negative flux in this area decreases. Ultimately, the negative flux in 
the spot itself should increase, but we cannot confirm this because the spot 
itself is outside our field-of-view.


\section{Conclusions}

We observe blueshifts (upflows) in the helium velocities along the central part 
of the fibrils, which hints at rising flux tubes. At their summits, matter can 
move only along the field line, and thus the matter has to rise with the flux 
tube. At the footpoints of the flux tubes we see a downward motion. These results 
are in qualitative agreement with previous results obtained by \cite{cit:bruzek69},
\cite{cit:saminat}, \cite{cit:spadaro}, \cite{cit:lagg07} and \cite{cit:xu10}.
Since we observe redshifts at both ends of the fibrils, we can 
exclude a siphon flow, in that case one would expect a blueshift on one side.
In total, 
this picture is in good agreement with the sketch of \cite{cit:bruzek69}. The 
rising flux tube does not gain much matter by condensations out of the 
surroundings, and after a while, not enough matter remains to maintain the 
downward drain and to absorb light in the helium line. The fibrils are no 
longer visible, and the downflow ends. This process is repeated when a new 
flux tube emerges.

The analysis of the helium and silicon data is ongoing work, and we have to 
refer to a future work answering the question: How is the magnetic field 
oriented in the layers of the fibrils? The analysis of the helium and silicon data 
will shed more light on the connection between the deep photosphere to the chromosphere.
In addition, we plan new observations to 
elucidate the physical properties of AFSs.


\acknowledgements We wish to thank the unknown referee for various 
comments that helped us to improve the manuscript significantly. 
The 1.5-meter GREGOR solar telescope was built by a German 
consortium under the leadership of the Kiepenheuer-Institut f\"ur Sonnenphysik 
in Freiburg (KIS) with the Leibniz-Institut f\"ur Astrophysik Potsdam (AIP), the 
Institut f\"ur Astrophysik G\"ottingen (IAG), the Max-Planck-Institut f\"ur 
Sonnensystemforschung in G\"ottingen (MPS), and the Instituto de 
Astrof{\'\i}sica de Canarias (IAC), and with contributions by the Astronomical 
Institute of the Academy of Sciences of the Czech Republic (ASCR). SJGM is 
grateful for financial support from the Leibniz Graduate School for Quantitative 
Spectroscopy in Astrophysics, a joint project of AIP and the Institute of 
Physics and Astronomy of the University of Potsdam. This work was supported by a 
program  of the Deutscher Akademischer Austauschdienst (DAAD) and the Slovak 
Academy of Sciences for project related personnel exchange (project 
No.~57065721). This work received additional support by project VEGA 2/0004/16. 
The GREGOR observations were obtained within the SOLARNET Transnational Access 
and Service (TAS) program, which is supported by the European Commission's FP7 
Capacities Program under grant agreement No.~312495.
SDO HMI data are provided by the Joint Science Operations Center -- Science data 
Processing.


\appendix

\section{Differential image rotation compensation} \label{sec:dirc}

The alt-azimuthal mount of the GREGOR-telescope \citep{cit:volkmer} causes an 
image rotation, which is not negligible even during a scan of just 18\,min. In 
the future, this image rotation can be compensated by a derotator, which was not 
available during our observations. To compensate for the differential image 
rotation during a scan, an IDL routine was developed. We assume that GAOS is 
locking on a structure next to the center of the scanned area keeping the solar 
structure at this position. Thus, the apparent axis of rotation falls 
together with this position, and we leave the central slit position untouched. 
According to the telescope elevation and azimuth, we calculate the relative 
rotation angle for each scan position with respect to the central one. The 
ephemeris are calculated at five-minute intervals, and then these results are 
interpolated to the exact time of each scan step. Next, we calculate the true 
position of each observed pixel within the map. Finally, we interpolate the map 
to a grid, which is equally spaced in seconds of arc along the two directions. 
Since we keep the orientation of the central slit position, a rotation of the 
whole image is still needed to orient the map with respect to the solar 
North-South direction. Thus, a rotation by a quite large angle might be 
required, while the differential rotation adds up to only a few degrees. 
Therefore, we separate these two steps. This routine will be included in the 
sTools IDL software library, which AIP's solar physics research group develops 
in the framework of the EU-funded SOLARNET project for ``High-Resolution Solar 
Physics''. For the present work, we apply this routine to the results of the 
SIR inversion but not to the spectra to save computation time.

The source code can be obtained on request from the corresponding 
author.


\begin{thebibliography}{}
  \bibitem[{{Berkefeld} {et~al.}(2012){Berkefeld}, {Schmidt}, {Soltau}, {von 
             der L{\"u}he},
             \& {Heidecke}}] {cit:gaos} Berkefeld, T., Schmidt, D., Soltau, D.,
             et al. 2012, \an, 333, 863
  \bibitem[{{Bommier} {et~al.}(2005){Bommier}, {Rayrole}, \& {Eff-Darwich}}]{cit:bommier}
             Bommier, V., Rayrole, J., \& Eff-Darwich, A. 2005, \aap, 435, 1115  
  \bibitem[{{Bruzek}(1967){Bruzek}}] {cit:bruzek67} Bruzek, A. 1967, \solphys, 2, 451  
  \bibitem[{{Bruzek}(1969){Bruzek}}] {cit:bruzek69} Bruzek, A. 1969, \solphys, 8, 29
  \bibitem[{{Collados}(1999){Collados}}]{cit:mcv} Collados, M.: 1999, in: 
             Third Advances in Solar Physics Euroconference: Magnetic Fields and Oscillations,
             eds. B. Schmieder, A. Hofmann, \& J. Staude, ASPC, 184, 3 
  \bibitem[{{Collados} {\&} {Schlichenmaier}(2002){Collados} \& {schlichenmaier}}]
          {cit:colschlich} {Collados}, M. \& {Schlichenmaier}, R. 2002, \aap, 381, 668
  \bibitem[{{Collados} {et~al.}(2007){Collados}, {Lagg}, {D{\'\i}az Garc{\'\i}a}, 
             {Hern\'andez Su\'arez}, {L\'Opez L\'opez}, {P\'aez Ma\~na}, \& {Solanki}}]
             {cit:tip2}{Collados}, M., {Lagg}, A., {D{\'\i}az Garc{\'\i}a}, J.J. et ~ al.
             2007, in: The Physics of Chromospheric Plasmas, eds. P. Heinzel, I. Dorotovi{\v c},
             \& Rutten, R.J., ASPC, 368, 611
  \bibitem[{{Collados} {et al.}(2012){Collados}, {L{\'o}pez}, {Pa{\'e}z}, {Hern{\'a}ndez}, 
            {Reyes}, {Calcines}, {Ballesteros}, {Diaz}, {Denker}, {Lagg}, {Schlichenmaier},
            {Schmidt}, {Solanki}, {Strassmeier}, {von der L{\"u}he}, \& {Volkmer}}]
            {cit:gris} Collados, M., L\'opez, R., P\'aez, E., et al. 2012, \an, 333, 872 
  \bibitem[{{Denker} {et~al.}(2012){Denker}, {von~der~L\"uhe}, {Feller}, {Arlt}, {Balthasar}, 
            {Bauer},{Bello Gonz{\'a}lez}, {Berkefeld}, {Caligari}, {Collados}, {Fischer},
            {Granzer}, {Hahn}, {Halbgewachs}, {Heidecke}, {Hofmann}, {Kentischer},
            {Klva{\v n}a}, {Kneer}, {Lagg}, {Nicklas}, {Popow}, {Puschmann}, {Rendtel},
            {Schmidt}, {Schmidt}, {Sobotka}, {Solanki}, {Soltau}, {Staude}, {Strassmeier},
            {Volkmer}, {Waldmann}, {Wiehr}, {Wittmann}, \& {Woche}}]{cit:denker12}
            Denker, C., von~der~L\"uhe, O., Feller, A., et~al. 2012, \an, 333, 810
  \bibitem[{{Gonz{\'a}lez Manrique} {et~al.}(2016){Gonz{\'a}lez Manrique}, {Kuckein},
            {Pastor Yabar}, {Collados}, {Denker}, {Fischer}, {G\"om\"ory}, {Diercke},
            {Bello Gonz{\'a}lez}, {Schlichenmaier}, {Balthasar}, {Berkefeld},
            {Feller}, {Hoch}, {Hofmann}, {Kneer}, {Lagg}, {Nicklas}, {Orozco Su\'arez},
            {Schmidt}, {Schmidt}, {Sigwarth}, {Sobotka}, {Solanki}, {Soltau}, {Staude},
            {Strassmeier}, {Verma}, {Volkmer}, {von der L{\"u}he}, \& {Waldmann}}]
            {cit:sergio} Gonz{\'a}lez Manrique, S.J., Kuckein, C., Pastor Yabar, A., 
            et~al. 2016, \an\ (this volume)
  \bibitem[{{Grigor'eva} {et al.}(2012){Grigor'eva}, {Shakhovskaya}, {Livshits}, \& 
           {Knyazeva}}]{cit:grigoreva} Grigor'eva, I., Yu., Shakhovskaya, A.N., 
            Livshits, M.A.,
           \& Knyazeva, I.S. 2012, Astron.\ Rep., 56, 887
  \bibitem[{{Hofmann} {et~al.}(2012){Hofmann}, {Arlt}, {Balthasar}, {Bauer}, {Bittner},
            {Paschke}, {Popow}, {Rendtel}, {Soltau}, \& {Waldmann}}]{cit:gpu} 
            Hofmann, A., Arlt, K., Balthasar, H., et al. 2012, \an, 333, 854
  \bibitem[{{Howard} {\&} {Harvey}(1964){Howard} \& {Harvey}}]{cit:howardharvey}
            Howard, R. \& Harvey, J. 1964, ApJ, 139, 1328
  \bibitem[{{Kneer}(2012){Kneer}}]{cit:kneer}
            Kneer, F. 2012, AN, 333, 790
  \bibitem[{{Kuckein} {et~al.}(2012a){Kuckein}, {Mart{\'\i}nez Pillet}, \& {Centeno}}]
            {cit:kuck12mag} Kuckein, C., Mart{\'\i}nez Pillet, V., \& Centeno, R. 2012a, 
            \aap, 539, A131
  \bibitem[{{Kuckein} {et~al.}(2012b){Kuckein}, {Mart{\'\i}nez Pillet}, \& {Centeno}}]
            {cit:kuck12vel} Kuckein, C., Mart{\'\i}nez Pillet, V., \& Centeno, R. 2012b, 
            \aap, 542, A112
  \bibitem[{{Lagg} {et~al.}(2007){Lagg}, {Woch}, {Solanki}, \&  {Krupp}}]{cit:lagg07}
            Lagg, A., Woch, J., Solanki, S.K., \& Krupp, N. 2007, \aap, 462, 1147
  \bibitem[{{Lagg} {et~al.}(2016) {Lagg}, {Solanki}, {Doerr}, {Mart{\'\i}nez Gonz{\'a}lez}, 
            {Rietm\"uller}, {Collados Vera}, {Schlichenmaier}, {Orozco Su{\'a}rez}, 
            {Franz}, {Feller}, {Kuckein}, {Schmidt}, {Asensio Ramos}, {Pastor Yabar},
            {von der L{\"u}he}, {Denker}, {Balthasar}, {Volkmer}, {Staude}, {Hofmann}, {Strassmeier}, 
            {Kneer}, {Waldmann}, {Borrero}, {Sobotka}, {Verma}, {Louis}, {Rezaei}, {Soltau},
            {Berkefeld}, {Sigwarth},{Schmidt}, {Kiess}, \& {Nicklas}}]{cit:lagg16}
            Lagg, A., Solanki, S.K., Doerr, H.-P., et~al. 2016, \aap, \textit{accepted};
            ArXiv 1605.06324
  \bibitem[{{Ma} {et~al.}(2015){Ma}, {Zhou}, {Zhou}, \& {Zhang}}]{cit:ma}
            Ma, L., Zhou, W., Zhou, G. \& Zhang, J. 2015, \aap, 583, A110
  \bibitem[{{Martres} {et~al.}(1966){Martres}, {Michard}, \& {Soru-Iscovici}}]{cit:martres66}
            Martres, M.-J., Michard, R., \& Soru-Iscovici, I. 1966, 
            Ann.\ Astrophys., 29, 249
  \bibitem[{{Ruiz Cobo} {\&} {del Toro Iniesta}(1992){Ruiz Cobo} \& {del Toro Iniesta}}]
            {cit:sir} Ruiz Cobo, B. \& del Toro Iniesta, J.C. 1992, 
            \apj, 398, 375
  \bibitem[{{Sasso} {et~al.}(2014){Sasso}, {Lagg}, \& {Solanki}}]{cit:sasso2014}
            Sasso, C., Lagg, A., \& Solanki, S.K. 2014, \aap, 561, A98
  \bibitem[{{Schad} {et~al.}(2013){Schad}, {Penn}, \& {Lin}}]{cit:schad}
           {Schad}, T.A., {Penn}, M., \& {Lin}, H. 2013, \apj, 768, 111    

  \bibitem[{{Schlichenmaier} {et~al.}(2011){Schlichenmaier}, {Bello Gonz{\'a}lez}, \& 
           {Rezaei}}]{cit:schliche273} 
           Schlichenmaier, R., Bello Gonz{\'a}lez, N., \& Rezaei, R. 2011,
           in: The Physics of Sun and Star Spots, eds. D.P. Choudhary \& K.G. Strassmeier, Proc.
           IAU Symposium, 273, 134           
  \bibitem[{{Schlichenmaier} {et~al.}(2012){Schlichenmaier}, {Rezaei}, \& {Bello Gonz{\'a}lez}}]
           {cit:schliche_hinode} 
           Schlichenmaier, R., Rezaei, R., \&  Bello Gonz{\'a}lez, N. 2012,
           in: 4$^{th}$ Hinode Science Meeting: Unresolved Problems and Recent Insights, 
           eds. L.R. Bellot Rubio, F. Reale, \& M. Carlsson, ASPC, 455, 61          
 \bibitem[{{Schmidt} {et~al.}(2012){Schmidt}, {von der L{\"u}he}, {Volkmer}, {Denker}, {Solanki},
            {Balthasar}, {Bello Gonz{\'a}lez}, {Berkefeld}, {Collados}, {Fischer}, {Halbgewachs},
            {Heidecke}, {Hofmann}, {Kneer}, {Nicklas}, {Popow}, {Puschmann}, {Schmidt}, {Sigwarth},
            {Sobotka}, {Soltau}, {Staude}, {Strassmeier}, \& {Waldmann}}]{cit:gregor}
            Schmidt, W., von der L\"uhe, O., Volkmer, R., et al. 2012, \an, 333, 796
  \bibitem[{{Schwartz} {et~al.}(2016){Schwartz}, {Balthasar}, {Kuckein},
            {Koza}, {G{\"o}m{\"o}ry}, {Ryb{\'a}k}, {Heinzel}, \& {Ku{\v c}era}}]{cit:schwartz2016}
            {Schwartz}, P., {Balthasar}, H., {Kuckein}, C., {et~al.} 2016, \an\ 
            (this volume)
  \bibitem[{{Solanki} {et~al.}(2003){Solanki}, {Lagg}, {Woch}, {Krupp}, \& {Collados}}]
            {cit:saminat} Solanki, S.K., Lagg, A., Woch, J., Krupp, N., \& Collados,M.
            2003, Nature, 425, 692
  \bibitem[{{Spadaro} {et~al.}(2004){Spadaro}, {Billotta}, {Contarino}, {Romano}, \& {Zuccarello}}]
            {cit:spadaro} Spadaro, D., Billotta, S., Contarino, L., Romano, P., \& Zuccarello, F.
              2004, \aap, 425, 309  
  \bibitem[{{Vargas Dom{\'\i}nguez} {et al.}(2012) {Vargas Dom{\'\i}nguez},{van Driel Gestelyi} \&
            {Bellot Rubio}}]{cit:vargas} {Vargas Dom{\'\i}nguez}, S., {van Driel Gestelyi}, L. 
            \& {Bellot Rubio}, L.R. 2012, \solphys, 278, 99
  \bibitem[{{Volkmer} {et~al.}(2012){Volkmer}, {Eisentr\"ager}, {Emde}, {Fischer}, {von~der~L\"uhe},
            {Nicklas}, {Soltau}, {Schmidt}, \& {Weis}}]{cit:volkmer}
            Volkmer, R., Eisentr\"ager, P., Emde, P., et~al. 2012, \an, 333, 816
  \bibitem[{{Xu} {et~al.}(2010){Xu}, {Lagg}, \& {Solanki}}]{cit:xu10}
             Xu, Z., Lagg, A., \& Solanki, S.K. 2010, \aap, 520, A77
\end{thebibliography}
\end{document}